\documentclass[11pt]{article}

\usepackage{scicite}
\usepackage[nospace,noadjust]{cite}

\usepackage{times}

\usepackage{graphics,graphicx}

\def\gtrsim{\lower 2pt \hbox{$\, \buildrel {\scriptstyle >}\over
{\scriptstyle \sim}\,$}}
\def\lesssim{\lower 2pt \hbox{$\, \buildrel {\scriptstyle <}\over
{\scriptstyle \sim}\,$}}
\def\ka{K$\alpha$}
\def\xs{Sgr~A*}

\def\chandra{{\sl Chandra}}

\def\apj{\rm{ApJ}}                 
\def\mnras{\rm{MNRAS}}             
\def\aap{\rm{A\&A}}                
\def\apjl{\rm{ApJ}}                
\def\nat{\rm{Nature}}              

\newenvironment{sciabstract}{%
\begin{quote} \bf}
{\end{quote}}

\newcounter{lastnote}

\title{Dissecting X-Ray-Emitting Gas Around the Center of Our Galaxy}

\author{Q. D. Wang,$^{1,2\ast}$ M. A. Nowak$^{3}$,  S. B. Markoff$^{4}$,
    F. K. Baganoff$^{3}$,\\ 
S. Nayakshin$^{5}$, F. Yuan$^{6}$,
J. Cuadra$^{7}$,   J. Davis$^{3}$, J. Dexter$^{8}$,\\
 A. C. Fabian$^{1}$, N. Grosso$^{9}$, 	D. Haggard$^{10}$, J. Houck$^{3}$,
L. Ji$^{11}$, Z. Li$^{12}$,\\ 
J. Neilsen$^{13,3}$, D. Porquet$^{9}$, F. Ripple$^{2}$,
R. V. Shcherbakov$^{14}$ \\
\normalsize{$^{1}$Institute of Astronomy, University of Cambridge, Madingley Road, Cambridge CB3 0HA, UK.}\\
\normalsize{$^{2}$Department of Astronomy, University of
  Massachusetts, Amherst, MA 01003, USA.}\\
\normalsize{$^{3}$Kavli Institute for Astrophysics and Space
  Research,}\\ 
\normalsize{Massachusetts Institute of Technology, Cambridge MA 02139, USA.}\\
\normalsize{$^{4}$Astronomical Institute, ``Anton Pannekoek'',
 University of Amsterdam,}\\
\normalsize{ Postbus 94249, 1090 GE, Amsterdam, The Netherlands.} \\
\normalsize{$^{5}$Department of Physics and Astronomy, University of
  Leicester, Leicester LE1 7RH, UK.}\\
\normalsize{$^{6}$Shanghai Astronomical Observatory, CAS, 80 Nandan Road, Shanghai 200030, China }\\ 
\normalsize{$^{7}$ Instituto de Astrof\'{i}sica, Pontificia Universidad Cat\'{o}lica de Chile, Chile.}\\
\normalsize{$^{8}$Theoretical Astrophysics Center and Department of Astronomy,}\\
\normalsize{ University of California, Berkeley, CA 94720-3411, USA}\\
\normalsize{$^{9}$Observatoire Astronomique de Strasbourg, Universit\'e
de Strasbourg,}\\
\normalsize{ CNRS, UMR 7550, Strasbourg, France}\\
\normalsize{$^{10}$CIERA Fellow, Center for Interdisciplinary
  Exploration and Research in Astrophysics,}\\
\normalsize{ Department of Physics and Astronomy, Northwestern University, Evanston, IL 60208, USA}\\
\normalsize{$^{11}$Purple Mountain Observatory, CAS, Nanjing, 210008, P.R. China.} \\
\normalsize{$^{12}$School of Astronomy and Space Science, Nanjing University, Nanjing, 210093, China.}\\
\normalsize{$^{13}$Einstein Fellow, Boston University, Boston, MA 02215, USA.}\\
\normalsize{$^{14}$Hubble Fellow; Department of Astronomy, University
  of Maryland,}
\normalsize{ College Park, MD 20742-2421, USA.}\\
\normalsize{$^\ast$To whom correspondence should be addressed; E-mail:  wqd@astro.umass.edu.}
}

\date{}
\clearpage
\eject

\begin{document} 




\maketitle 

\begin{sciabstract}
  Most supermassive black holes (SMBHs) are accreting at very low
  levels and are difficult to distinguish from the galaxy centers
  where they reside.  Our own Galaxy's SMBH provides a uniquely
  instructive exception, and we present a close-up view of its
  quiescent X-ray emission based on 3 mega-second of {\sl Chandra}
  observations. Although the X-ray emission is elongated and aligns well with a surrounding disk of massive stars, we can rule out a concentration of low-mass coronally active stars as the origin of the emission based on the lack of predicted Fe K$\alpha$ emission. The extremely weak H-like Fe
  K$\alpha$ line further suggests the presence of an outflow from the
  accretion flow onto the SMBH.   These results provide important
  constraints for models of the prevalent radiatively inefficient
  accretion state. 
\end{sciabstract}

The nucleus of our Galaxy offers a multitude of unique opportunities
for observing the interplay between a SMBH and its immediate
surroundings.  The SMBH, named Sgr A*, has a mass of $ 4.1 \times 10^6
M_\odot$~\cite{Ghez2008,Gillessen2009}, and at its distance of 8
kpc~\cite{Ghez2008}, an arcsecond (1$^{\prime\prime}$) subtends $1.2
\times 10^{17}$ cm, or $1.0 \times 10^5 r_s$, where $r_s = 1.2 \times
10^{12} {\rm~cm}$ is the Schwarzschild radius.  The X-ray
emission of \xs\ typically has an unabsorbed 2-10~keV luminosity
($L_x$) of a few times $ 10^{33}
{\rm~erg~s^{-1}}$~\cite{Baganoff2003}, or a factor of $\sim 10^{11}$
lower than the canonical maximum allowed (Eddington) luminosity of the
SMBH, representing a common ``inactive'' state of galactic nuclei in
the Local Universe. Because of its proximity, \xs\ allows us to study
this extremely low-$L_X$ state in unparalleled detail
[e.g.,~\cite{Genzel2010,Markoff2010}]. 

It is believed that \xs\ feeds off the winds from surrounding massive
stars~\cite{Coker1997,Quataert1999,Cuadra2008}.  At the so-called
Bondi capture radius $r_B \sim 4^{\prime\prime}
(T_a/10^7~\rm~K)^{-1}$~\cite{Bondi1952}, the gravitational pull of the
SMBH wins over the expanding motion of the medium with effective
temperature $T_a$. The corresponding Bondi capture rate is estimated
to be $\dot{M}_B \sim 1 \times 10^{-5} M_\odot {\rm~yr^{-1}}$ [e.g.,
\cite{Baganoff2003}]. If \xs\ indeed accretes at this rate and if the
10\% X-ray emission efficiency of a ``normal" active galactic nucleus
applies, one would predict a luminosity of $L_x \sim 10^{41}
{\rm~erg~s^{-1}}$.  That the observed $L_x$ is nearly a factor of
$\sim 10^8$ smaller has led to a renaissance of radiatively
inefficient accretion flow (RIAF) models~\cite{Shapiro1976,Rees1982}, including self-similar
solutions~\cite{Narayan1994,Narayan1995,Quataert2000,Blandford1999,Blandford2004,Begelman2012} and numerous hydrodynamic or magneto-hydrodynamic simulations
of various complexities and dynamic ranges [e.g.,~\cite{Yuan2012a,Yuan2012b,Li2013,Narayan2012,Dibi2012,Drappeau2013}]. Many of the recent model development
were stimulated by sub-mm polarization/Faraday rotation measurements,
providing stringent constraints on the accretion rate in the innermost
region ($r \lesssim 10^2 r_s$) of \xs [e.g.,~\cite{Marrone2007}]. But
controversies remain as to which model (if any) may apply
[e.g.,~\cite{Narayan 2012,Yuan2012b}].

High-angular resolution observations of Sgr A* -- such as those
provided by the \chandra\ X-ray Observatory --- can in principle probe
the accretion phenomenon not only in its innermost regions, but also
the outer boundary conditions at flow onset.  However, there are
substantial uncertainties regarding the interpretation of the
quiescent emission of \xs\, which previous \chandra\ observations
showed to be extended with an intrinsic size of about
$1.4^{\prime\prime}$~\cite{Baganoff2003}. 
How that emission is distributed within this region, and how much is due to a bound
accretion flow versus other components, remain unclear.  Furthermore, it has been
proposed~\cite{Baganoff2003,Sazonov2012} that a substantial or even
dominant fraction of the emission could arise from a centrally-peaked
population of coronally active, low-mass main-sequence stars around
\xs, which is allowed by current near-infrared observations. This
scenario predicts a 6.4~keV emission line with the equivalent width
(EW) in the range of $50-100$~eV, due to fluorescence of photospheric
weakly ionized irons, irradiated by coronal flare X-rays.  It is thus
essential to test this hypothesis before we can assign the X-ray
emission to the accretion flow onto the SMBH.

We use data taken during the Sgr A* X-ray Visionary
Program [XVP;~\cite{zzz}]. The ACIS-S (Advanced CCD Imaging
Spectrometer-Spectroscopy), combined with the HETG  (High-energy Transmission
Grating), was placed at the aim point of the telescope. 
The on-axis spatial resolution of \chandra\ is $\sim 0.4^{\prime\prime}$ (FWHM), whereas 
the spectral resolution of the 0th-order ACIS-S/HETG data is about a factor of 
$\sim 2$ better than that of the previous ACIS-I observations and thus enables critical 
spectroscopic diagnostics of the X-ray emission.

\begin{figure*} 
\centering
\includegraphics[width=6.in,angle=0]{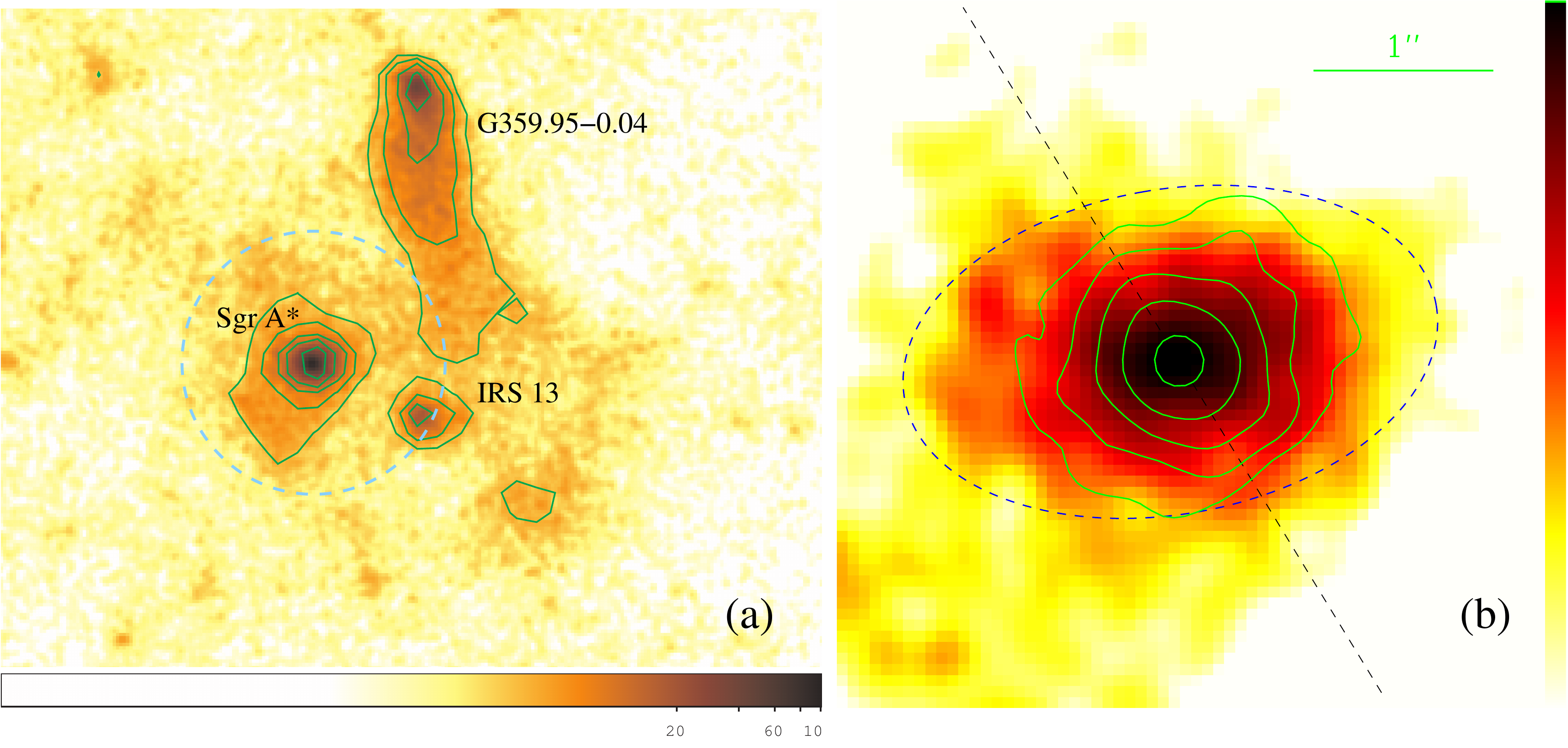}
\caption{
X-ray images of \xs\ in quiescence. (a) An image
constructed with the XVP 0th-order ACIS-S/HETG data in the
1-9~keV band.
The contours are at 1.3, 2.2, 3.7, 6.3, and  11
$\times 10^{-4} {\rm~cts~s^{-1}~arcsec^{-2}}$. North is up and East is
to the left. 
The dashed circle around \xs\ marks its Bondi capture
radius (assumed to be 4$^{\prime\prime}$).
(b) A close-up of \xs. The emission is decomposed into extended (color image) and 
point-like (contour) components. The latter component is modeled with
the net flare emission~\cite{zzz} and is illustrated as the intensity
contours at 
0.3, 0.6, 1.2, 2.4, and 5 
counts per pixel. The straight dashed line marks the orientation of
the Galactic plane, whereas the dashed ellipse of a 1.5$^{\prime\prime}$ semi-major axis
illustrates the elongation of the primary massive stellar disk, which has an 
inclination of $i \sim 127^\circ$, a line-of-nodes position angle of 
$100^\circ$ (East from North), and a radial density distribution
$\propto r^{-2}$ with a sharp inner cut off at $r \approx  1^{\prime\prime}$~\cite{Paumard2006}.
}
\label{f:fig1}
\end{figure*}

These observations reveal X-ray emission from \xs\ that appears
substantially more extended than a point-like source
[Fig.~\ref{f:fig1}a;~\cite{zzz}]. After subtracting a point-like
contribution, which accounts for 20\% of the total quiescent
emission~\cite{zzz}, an east-west-elongated X-ray enhancement emerges
around \xs\ on $\sim 1^{\prime\prime}$-1.5$^{\prime\prime}$ scales
(Fig.~\ref{f:fig1}b).  This relatively symmetric enhancement
morphologically resembles the so-called clockwise stellar disk, which
is the most notable kinematically organized structure known for the
massive stars around \xs~\cite{Paumard2006,Lu2009}. 
Irregular low-surface-brightness features (e.g., spurs toward the north-east and
south-east; Fig.~\ref{f:fig1}b) appear on scales greater than the
enhancement, but still roughly within the Bondi radius. These features
are softer (more prominent in the 1-4~keV band than in the 4-9~keV
band) than the smoothly-distributed background.

The quiescent X-ray emission is also spectrally distinct from the
point-like flare emission~\cite{zzz}, which is considerably harder.
The accumulated flare spectrum can be well characterized by a power
law with photon index 2.6(2.2, 3.0), and a foreground absorption with 
an equivalent hydrogen 
column density of $N_H = 16.6(14.1, 19.4) \times 10^{22} {\rm~cm^{-2}}$
[see~\cite{note} for the definition of the uncertainties], consistent
with the previous analysis of individual bright
flares~\cite{Porquet2008,Nowak2012}. Because of the simplicity of the
flare spectrum, this $N_H$ measurement can be considered a reliable
estimate of the foreground X-ray-absorbing column density along the
sightline toward \xs, and hence a useful constraint in modeling the
more complex spectrum of the quiescent X-ray emission.

\begin{figure*} 
\includegraphics[width=6.5in]{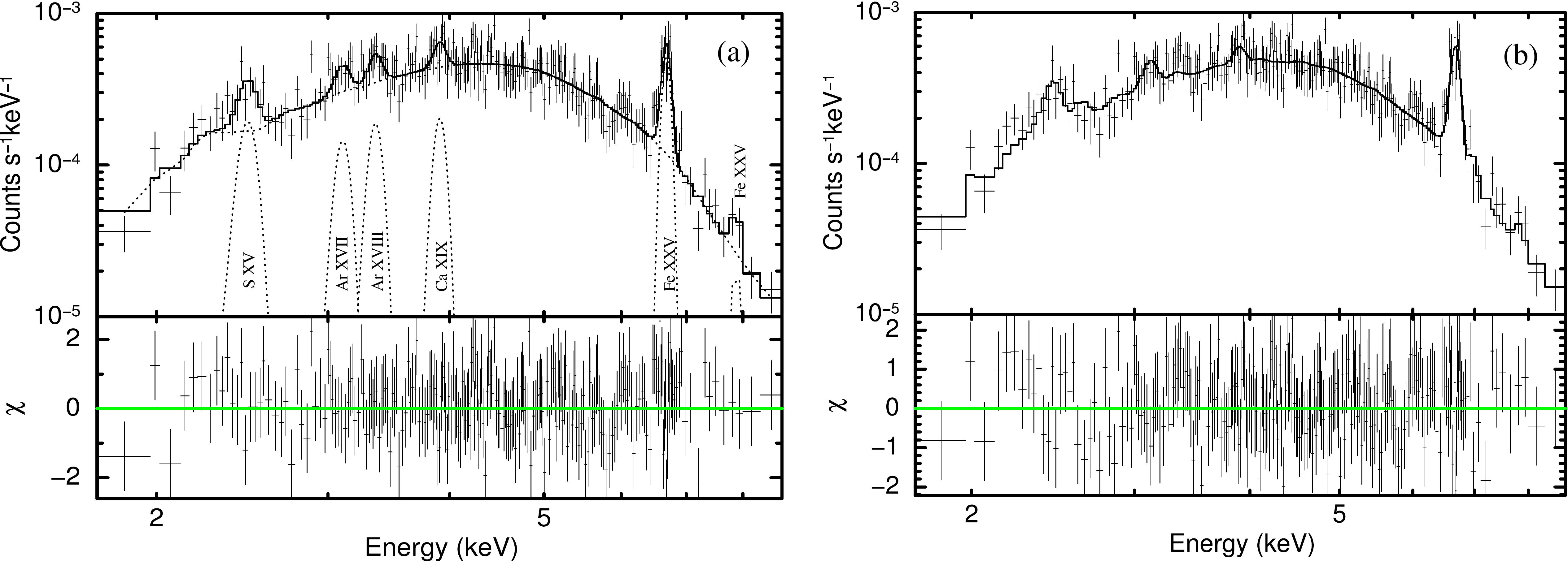}
\caption{
The X-ray spectrum of \xs\ in quiescence [extracted from the inner circle of 1.5$^{\prime\prime}$
radius and local-background-subtracted;~\cite{zzz}] and model fits: (a)
zero-metallicity 1-T thermal bremsstrahlung continuum plus various
Gaussian lines (Table~\ref{t:line_para});
(b) {\it  RIAF} model with the best-fit $\gamma=2s/\theta=1.9$, where $s$ and $\theta$ are the indices parametrizing the density and
temperature profiles, respectively~\cite{zzz}. }
\label{f:fig2}
\end{figure*}

By contrast, the quiescent spectrum shows prominent emission lines
(Fig.~\ref{f:fig2}). In addition to the previously known Fe K$\alpha$
emission at $\sim 6.7$~keV, K$\alpha$ lines of several other species
(He- and H-like Ar, He-like S and Ca), as well as He-like Fe K$\beta$
are also apparent. We find that a single-temperature (1-T) plasma with
metal abundance set equal to zero describes the overall (continuum)
shape of the observed spectrum well, while individual Gaussians can be
used to characterize the centroid, flux, and EW of six most
significant emission lines [Fig.~\ref{f:fig2}a; \cite{zzz}]. There is
little sign of the diagnostic K$\alpha$ emission lines from
neutral/weakly ionized Fe at $\sim 6.4$~keV or H-like Fe 
at 6.97~keV.

We can reject the stellar coronal hypothesis, based on our temporal,
spatial, and spectral results.  First, our spectrum does not confirm
the presence of the 6.4-keV line, previously suggested by
~\cite{Sazonov2012}. Measuring the 6.4-keV line, which is adjacent to
the strong highly ionized Fe \ka\ line, is difficult when the spectral
resolution is poor, as was the case for the previous ACIS-I spectrum,
because the measurement then depends sensitively on the assumed
thermal plasma model.  Our measured upper limit to the EW of the
6.4-keV line emission, 22~eV~\cite{zzz}, is more than a factor of 2
below the range predicted before~\cite{Sazonov2012}.  Second, the
quiescent emission shows no significant variation on time scales of
hours or days, as expected from the sporadic giant coronal flares of
individual stars.  Third, if the bulk of the quiescent emission is due
to stellar coronal activity, then it should also account for some of
the detected X-ray flares, especially the relatively long and weak
ones~\cite{Sazonov2012}. However, no sign of any line emission, even
the strong Fe \ka\ line, is found in any flare spectra [individual or
accumulated;~\cite{Neilsen2013,zzz}].  Fourth, the spatial
distribution of the weak flares is as point-like as the strong ones,
in contrast to the extended quiescent emission~\cite{zzz}. Therefore, we conclude that neither flare nor
quiescent emission originates from stellar coronal activity, although
the latter could still give a small ($\lesssim$25\%) contribution to the
quiescent emission.

As described above, the diffuse X-ray emission on the scale of
$\lesssim 1.5^{\prime\prime}$ (or $1.5 \times 10^5 r_s$)
morphologically resembles a gaseous disk with inclination angle
and line-of-nodes similar to those of the stellar disk around
\xs. This scale, a factor of $\gtrsim 2$ smaller than $r_B$,
corresponds to the sound-crossing distance over $\sim 10^2 (T_a/10^7
\rm~K)^{-1/2}$ years (where $T_a$ is the ambient gas temperature), which is about the time since the last major
burst of \xs, as inferred from X-ray light echoes
[e.g.,~\cite{Terrier2010,Nobukawa2011}]. The burst, making \xs\ a
factor of $\sim 10^6$ brighter than its present state over a period of
several $10^2$ years, could have substantially altered the
surroundings, as well as the accretion flow itself.  A stable
accretion flow would then need to be re-established gradually, roughly
at the sound-crossing speed.  This flow would be expected to carry the net angular
momentum of the captured wind material with an orientation similar to
that of the stars, which could explain the morphological
similarities. However, the recent hydrodynamic simulations of the
stellar wind accretion~\cite{Cuadra2008} suggest that a Keplerian
motion-dominated flow should occur on much smaller scales ($r \lesssim
0.1^{\prime\prime}$). If true, then the plasma on larger
scales must be supported mainly by large gradients of thermal,
magnetic, cosmic-ray, and/or turbulence pressures, and/or the motion must be strongly affected by the outward transport of angular momentum from the accretion flow, as indicated in
various simulations [e.g.,~\cite{Yuan2012b}].

One can further use the relative strengths of individual lines as
powerful diagnostics of the accretion
flow~\cite{Narayan1999,Perna2000}. X-ray emission lines trace the
emission measure distribution as a function of plasma temperature.
The ionic fraction of He-like S (S~{\sc XV}), for example, peaks at
$kT\sim 1.4$ keV, while the He-like Fe dominates in the temperature
range of $\sim 1.5 -7$ keV and the H-like Fe peaks sharply at $\sim 9
$ keV. At temperatures $\gtrsim 12$ keV, Fe becomes nearly fully
ionized. Thus if the radial dependence of the temperature can be modeled, the
density or mass distribution as a function of radius can be inferred.

The extremely weak H-like Fe \ka\ line in the \xs\ spectrum
immediately suggests a RIAF with a very low mass fraction of the
plasma at $kT \gtrsim 9$ keV, or a strong outflow at radii $r \gtrsim
10^4 r_s$~\cite{zzz}. Radially resolved analysis further shows that the X-ray
spectrum becomes increasingly hard with decreasing radius and that the
line centroid and EW of the Fe \ka\ line also changes with the
radius.  These characteristics can naturally be explained
by the presence of plasma at a relatively low temperature in the
region, consistent with the onset of the RIAF from captured stellar
wind material at $r \sim 10^5 r_s$.

As detailed in~\cite{zzz}, we can quantitatively check the consistency
of the X-ray spectrum of the quiescent \xs\ with various RIAF solutions.  In particular, we can reject (with a null hypothesis
probability of $10^{-6}$) a no-outflow solution (i.e., a constant mass
accretion rate along the accretion flow leading to a radial plasma
density profile $n \propto r^{-3/2+s}$, in which $s=0$).  This
solution substantially over-predicts the H-like Fe \ka\
line, while other lines are not fully accounted for. The predicted shape
of the model spectrum is also far too flat, resulting in a
substantially reduced $N_H [= 8.0 (7.6, 8.3) \times 10^{22}
{\rm~cm^{-2}}$], inconsistent with other estimates. Therefore, the
no-outflow solution can be firmly rejected, both statistically and
physically.

We find that a {\it RIAF} model with a flat radial density profile
(i.e., $s \sim 1$) provides an excellent fit to both the continuum and
lines data (Fig.~\ref{f:fig2}b), and gives estimates of both the
plasma metal abundance and the foreground absorption column consistent
with other independent measurements~\cite{zzz}. With this fit, we can
infer the radial emission structure of the accretion flow.  Although
the line emission is dominated by the outer, cooler region of the RIAF
($r \gtrsim 10^4 r_s$), the innermost hot component contributes
primarily to the continuum emission, mostly via bremsstrahlung
processes. We find that this latter contribution accounts for only
$\lesssim 20\%$ of the observed quiescent X-ray luminosity (consistent
with the spatially decomposed fraction), in stark contrast to the $s
=0$ solution,
whose X-ray emission is completely dominated by the innermost regions
($r \lesssim 10^2 r_s$). 

Further insight into the physical processes involved in the RIAF can
be obtained from comparing our X-ray measurements, most sensitive to
the outer radial density profile, with existing constraints on the
accretion properties in the innermost region. For example,
submillimeter Faraday rotation measurements~\cite{Marrone2007} set a
lower limit to the accretion rate of $2\times 10^{-9} M_\odot
{\rm~yr^{-1}}$ into the innermost region. Assuming that a large
fraction of the Bondi accretion rate, $\dot{M}_B \sim 1 \times 10^{-5}
M_\odot {\rm~yr^{-1}}$, initially ends up in the RIAF at a radius
$\sim 10^5 r_s$, then an $s \sim 1$ density profile can exist down to
radius $r_i \sim 10^2 r_s$. This $s \sim 1$ density profile over a
broad radial range is consistent with various {\it RIAF} solutions
[e.g.,~\cite{Quataert2000,Begelman2012}] and numerical simulations
[e.g., \cite{Yuan2012a}].  The Faraday rotation measurements also
yield an upper limit to the accretion rate of $2\times 10^{-7} M_\odot
{\rm~yr^{-1}}$, assuming that the magnetic field is near
equipartition, ordered, and largely radial in the
RIAF~\cite{Marrone2007}.
With this upper limit, assuming that $r_i \sim 10^2 r_s$ we can infer $s
> 0.6$ and hence $\theta \gtrsim 0.6$ for the radial temperature
profile $T \propto r^{-\theta}$ of the RIAF. This constraint on
$\theta$ places a fundamental limit on thermal conduction-induced heat
outflow~\cite{Shcherbakov2010}.

The flat density profile of the flow (i.e., $s \sim 1$) suggests the
presence of an outflow that nearly balances the inflow~\cite{zzz}. As
a result, $\lesssim 1\%$ of the matter initially captured by the SMBH
reaches the innermost region around \xs, limiting the accretion power
to $\lesssim 10^{39} {\rm~erg~s^{-1}}$. Much of this power is probably
used to drive the outflow, which could affect the environment of the
nuclear region and even beyond [e.g.,~\cite{Crocker 2012}]. This
combination of the low energy generation and high consumption rates of
\xs\ naturally explains its low bolometric luminosity (a few times
$10^{36} {\rm~erg~s^{-1}}$) and its lack of powerful jets.
Observations of nearby elliptical galaxies with jet-inflated cavities
in diffuse X-ray-emitting gas reveal a relation between the jet and
Bondi accretion powers (in units of $10^{43} {\rm~erg~s^{-1}}$),
$P_{jet} \approx 0.007 P_{Bondi}^{1.3}$, where $P_{Bondi}$ is assumed
to have a 10\% efficiency of the Bondi mass accretion
rate~\cite{Allen2006}. We speculate that the nonlinearity of this
relation is due to increasing outflow rates with decreasing Bondi
power, leading to a reduced energy generation rate in the innermost
region around a SMBH, where jets are launched.  Extrapolating the
relation down to the case for \xs, we may expect a jet to Bondi power
ratio of only $\sim 0.2\%$, consistent with the above inferred
fractional mass loss due to the outflow.

The progress in understanding the accretion flow characteristics and
orientation for our closest SMBH will lead to key constraints for the
modeling of such phenomena in general.  For \xs\ specifically these
results will impact models of the ``shadow'' of the SMBH
[e.g.,~\cite{Falcke2000,Moscibrodzka2009}], which can be measured by very long
baseline interferometry at (sub)millimeter
wavelengths~\cite{Doeleman2008}, and the radiation from hydrodynamic
interactions of orbiting stars [e.g., S2;~\cite{Nayakshin2005}] and
the G2 object with ambient media at distances down to $\sim 10^3 r_s$ from the SMBH
[e.g.,~\cite{Gillessen2013}]. 


\bibliographystyle{Science}

{\bf Acknowledgments:}
We thank the Sgr~A* \emph{Chandra} XVP Collaboration
(www.sgra-star.com/collaboration-members), which is supported by SAO grant
GO2-13110A, and are grateful to \emph{Chandra} Mission Planning for their
support during our 2012 campaign.  QDW thanks the hospitality of the
Institute of Astronomy and the award of a Raymond and Beverley Sackler
Distinguished Visitor fellowship.  We acknowledge the role of the Lorentz
Center, Leiden: the Netherlands Organization for Scientific Research Vidi
Fellowship (639.042.711; SM).  Further support from NASA was from
Hubble Fellowship grant HST-HF-51298.01 (RVS); from
Einstein Postdoctoral Fellowship grant PF2-130097 (JN); and from SAO contract
SV3-73016 to MIT for support of the \emph{Chandra} X-ray Center, which is
operated by SAO for and on behalf of NASA under contract NAS8-03060;
from CONICYT-Chile through FONDECYT (11100240),
Basal (PFB0609) and Anillo (ACT1101) grants (JC); and from 
NSFC and the 973 Project (2009CB82400) of
China (FY).  The X-ray data used here are available from the \emph{Chandra}
archive at asc.harvard.edu.

\clearpage

\clearpage
\setcounter{page}{1}

\title{\bf \large \centerline {Supplementary Materials for}}
\title{\bf \centerline{Dissecting X-ray-emitting Gas around  the
    Center of our Galaxy}}
\title{\centerline{Q. D. Wang, M. A. Nowak,  S. B. Markoff,
    F. K. Baganoff, S. Nayakshin, F. Yuan,
J. Cuadra,   J. Davis, J. Dexter, }}
\title{\centerline{A. C. Fabian, N. Grosso,
 	D. Haggard, J. Houck, L. Ji, Z. Li, J. Neilsen, D. Porquet,
F. Ripple, R. V. Shcherbakov}}

\title{\centerline{correspondence to:  wqd@astro.umass.edu}}
\setcounter{figure}{0}\renewcommand{\thefigure}{S.\arabic{figure}} 
\setcounter{figure}{0}\renewcommand{\thetable}{S.\arabic{table}} 

\noindent{\bf This PDF file includes:}

\noindent Materials and Methods\\
\noindent Figs. S1 to S3\\
\noindent Table S1 \\
\noindent References (45-56)

\noindent{\bf Materials and Methods}

The \chandra\ X-ray Visionary Project (XVP) to observe the Milky Way's
SMBH, Sgr A*, and the surrounding inner few arcminutes, culminated in 
38 observations for a total 3 mega-second exposure, taken
during February 6 to October 29, 2012. This \chandra\
Cycle 13 program utilized the High
Energy Transmission Gratings (HETG) in combination with the Advanced
CCD Imaging Spectrometer-Spectroscopy (ACIS-S) to accomplish high
spatial, spectral, and temporal resolutions. Only the 0th-order
ACIS-S/HETG data from the observations are used in the present
work. The spectral resolution of the data is about a factor of 
$\sim 2$ better than that of the previous Advanced CCD Imaging
Spectrometer-imaging (ACIS-I) observations 
(which suffered from a severe charge transfer inefficiency).

\noindent\underline{Data Reduction}

The data are reduced via standard CIAO processing routines (version
4.5; Calibration Database version 4.5.6). In particular, source
detection is carried out for individual observations and later for the merged data set. 
Sources detected within $3^{\prime}$ of 
\xs\ and with position errors less than $0.2^{\prime\prime}$ are  used 
by the routine {\sl reproject$\_$aspect} to match each observation to 
the longest one (ObsID 13842), which is aligned first 
to the radio position of \xs\  [J2000: $RA=17^h45^m40.041^s, 
Dec=-29^\circ 00^\prime 28.12^{\prime\prime}$]. 
The merged data can be treated approximately like a single
observation, because the differences among the pointing directions of
individual observations are all within a 14$^{\prime\prime}$ radius. 
Consistency checks for both spectral and spatial results have
also been carried out with existing ACIS-I observations. We find no
apparent calibration issues with the XVP ACIS-S/HETG data.

\noindent\underline{Decomposition of the \xs\ data in the time domain}

The data on \xs\ are divided into two parts: one in various
flare periods and the other in ``flare-free'' or  quiescent periods.
A census of flares detected in the data can be found in~\cite{Neilsen2013}. Different detection algorithms give slightly different
properties (e.g., peak fluxes), as well as different numbers of flares, mostly at the  low flux end.  
In this work, we have adopted the flare catalog from the detection with the ``Bayesian
Blocks'' routine [see Appendix of~\cite{Neilsen2013}]. The catalog contains 45
flares, a few more than the number obtained with a 
Gaussian kernel detection scheme. Therefore, our adopted flare catalog leads
to a relatively conservative definition of the 
quiescent X-ray emission. Various tests, such as auto-correlation,  suggest that the quiescent emission is nearly a
pure Poisson process of a constant photon flux; the upper limit to the
flare contribution
to the emission within $\lesssim 1.5^{\prime\prime}$ is 10\%, consistent with the very flat fluence
function of the detected flares~\cite{Neilsen2013}.  The total
exposure is 0.18 mega-seconds for the flares and 2.78 mega-seconds
for the quiescent emission. The data accumulated from the quiescent periods
(after exposure correction) can be subtracted from
those of the flare periods to obtain a net accumulated flare spectrum or image.

\noindent\underline{Flare image and spatial decomposition of the \xs\ quiescent X-ray
  emission}

\begin{figure*} 
\centering
\includegraphics[width=6.in,angle=0]{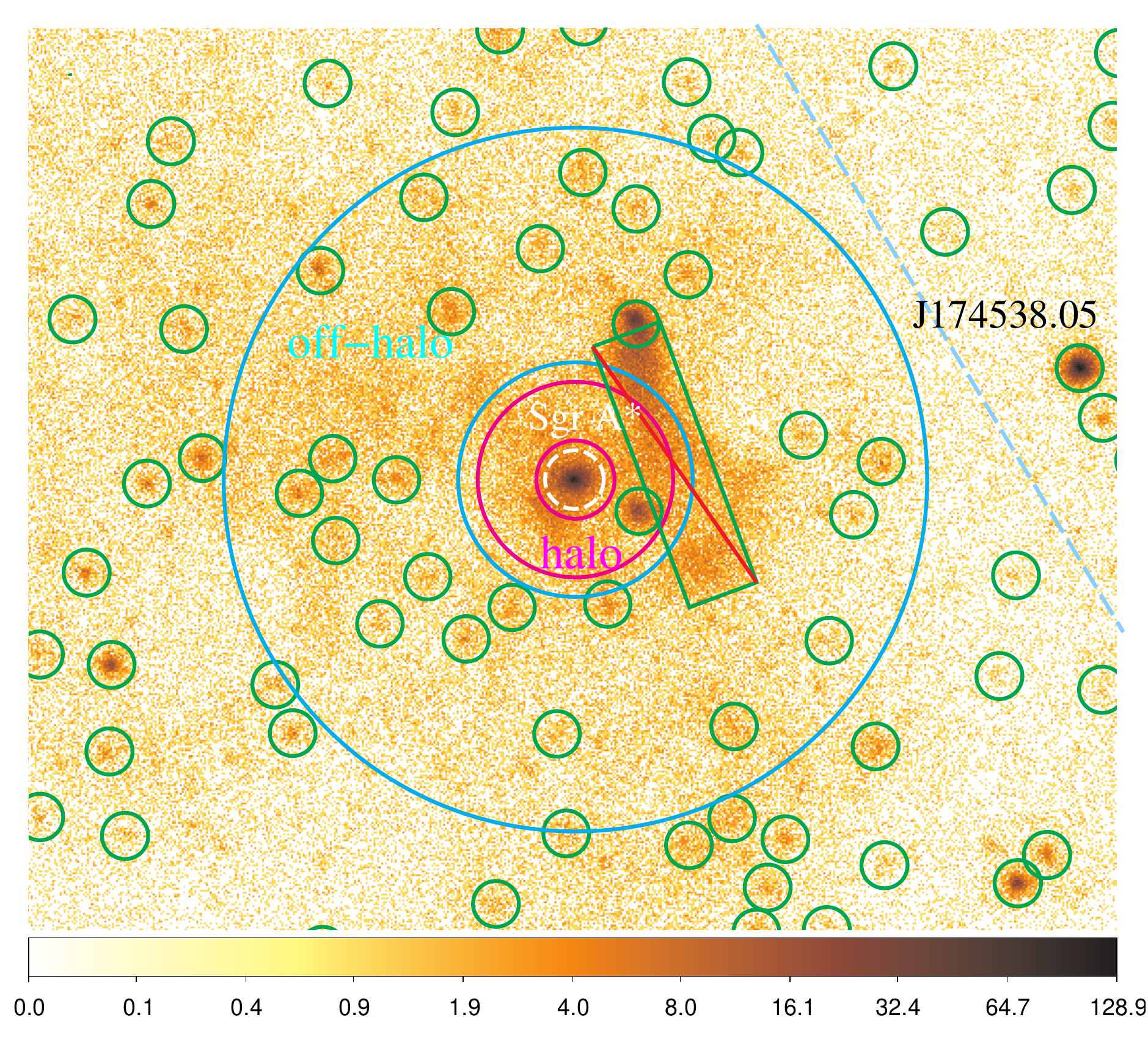}
\caption{
X-ray Overview  of the field ($56^{\prime\prime} \times
46^{\prime\prime}$) around \xs. The ACIS-S/HETG image is constructed
in the same way as Fig.~\ref{f:fig1}a. The small central (dashed white) circle 
marks the on-\xs\ spectral extraction region, while the two large
concentric (solid magenta and cyan) annuli outline the \xs-halo and
off-halo background
subtraction regions. The 
green box and small circles mark areas that are significantly contaminated by
an extended PWN and 
detected sources (1.5 times 70\% energy-encircled radii) and are thus excluded
from the spectral extractions. 
}
\label{f:sfig1}
\end{figure*}

\begin{figure} 
\centering
\includegraphics[width=7in,angle=0]{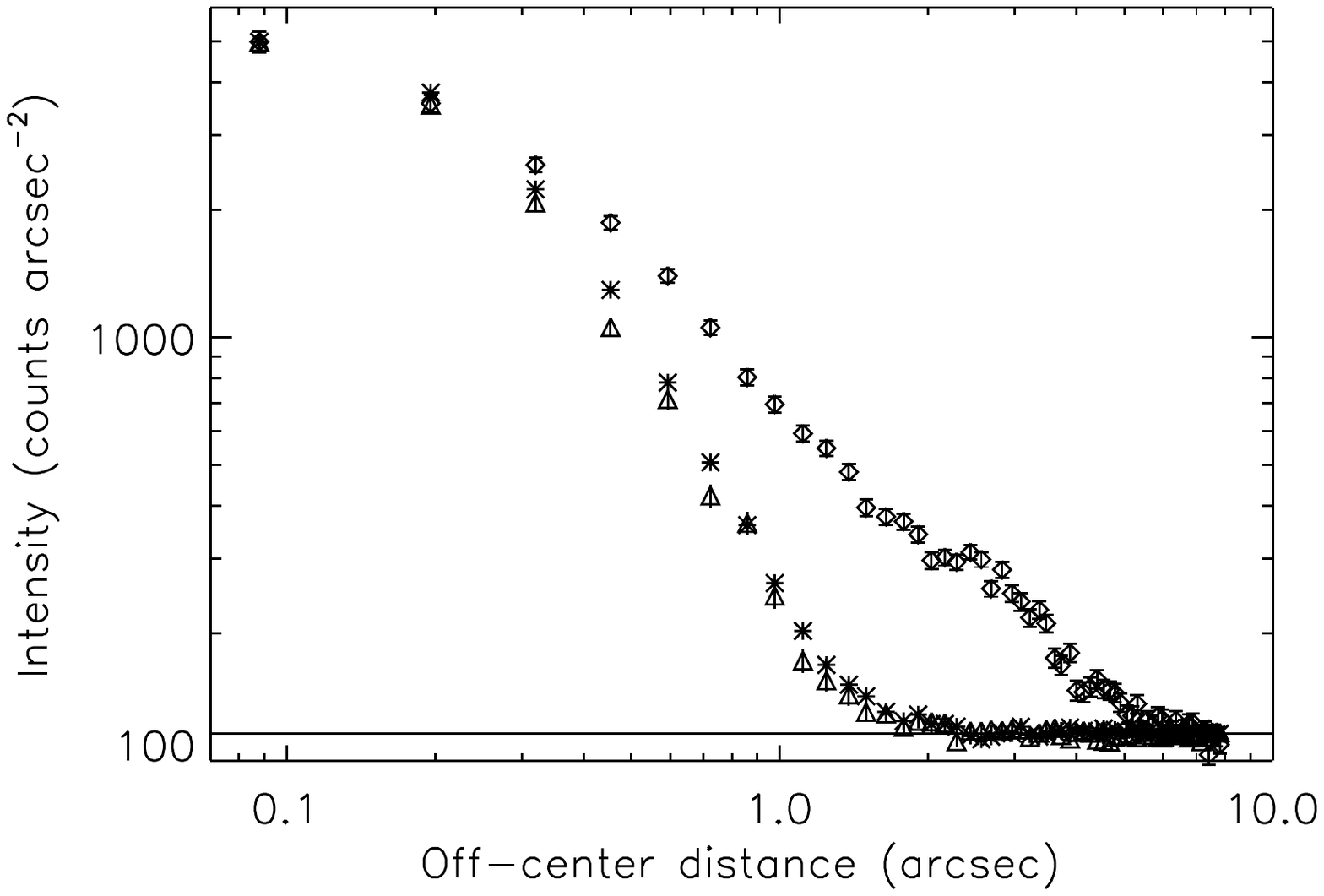}
\caption{
Comparison of the radial 1-9~keV intensity profiles of \xs\ in
quiescence ({\sl diamonds}) and in flares ({\sl triangles}, as well as 
J174538.05-290022.3 ({\sl crosses}). Both the
\xs\ flare and J174538.05-290022.3 profiles have been normalized to
the intensity of the first bin of the quiescent \xs\  profile. This
normalization is performed after subtracting the
corresponding local background of
each profile, estimated as the median value in the 6$^{\prime\prime}$-8$^{\prime\prime}$ range. The
profiles are then shifted to the local background level (the solid 
horizontal line) of the
quiescent \xs\ profile.
}
\label{f:sfig2}
\end{figure}

We construct an image of flares by stacking them and subtracting the 
exposure-corrected quiescent contribution. The three brightest ones
are excluded from this stacking to minimize potential pile-up effects [two or more photons in a single ``readout frame''
  being registered as a single event, or rejected as a non X-ray
  event;~\cite{Davis 2001}] The resultant net flare image  (represented by the contours in
Fig.~\ref{f:fig1}b) is consistent with the expected on-axis point-spread
function of the instrument and shows a round morphology 
(with statistical noise). The radial intensity
profile of the \xs\ flare emission is slightly narrower than that of
J174538.05-290022.3 (Fig.~\ref{f:sfig2}).
The profile of this latter source is included here as a reference of
a point-like distribution. The source shows  high variability of X-ray intensity with time
and is thus point-like. With a spectrum indicative of strong
absorption and projected only 27$^{\prime\prime}$ west of \xs\ (Fig.~\ref{f:sfig1}), 
the source is most likely located at the Galactic center or beyond.
 Therefore, the intensity profile of
the source can reasonably be used as an empirical reference of the
combined angular dispersion of X-ray radiation 
due to both the instrument point-spread function and the scattering by
dust along the line of sight. 

To better assess the extended morphology of the quiescent emission, we subtract from it the
flare image scaled to minimize the roundness of the resultant image,
but not to produce centrally depressed or peanut-shaped morphology.
The adopted scaling corresponds to a subtracted point-like component
accounting for 20\% of the total quiescent emission --- a fraction that is
also consistent with the spectral decomposition of the emission to be
described later.

\noindent\underline{Spectral analysis}

We adopt the same 1.5$^{\prime\prime}$ radius region 
as used previously~\cite{Baganoff2001,Xu2006,
Sazonov2012} to extract X-ray
spectra of \xs\  in its flare and quiescent states. We also extract a
``\xs-halo'' spectrum from a concentric annulus of the inner and outer
radii of 2$^{\prime\prime}$ and 5$^{\prime\prime}$; this outer 
bound is comparable to the Bondi radius
within its uncertainty. The \xs-halo spectrum is not only used as the local
background of the on-\xs\ spectrum, but analyzed to assess the ambient
X-ray properties. To do so, 
we further obtain an ``off-halo'' spectral background from a 
concentric 6$^{\prime\prime}$-18$^{\prime\prime}$ annulus (Fig.~\ref{f:sfig1}). The comet-shaped pulsar wind nebula G359.95-0.04
\cite{Wang2006}, as well as the detected
sources, are excluded from these annuli. Each on-source spectrum is
grouped to achieve a respective background-subtracted
signal-to-noise ratio of $\gtrsim 3$. The accurate background
subtraction is only moderately important for the quiescent \xs\
spectrum. It has a 1-9~keV count rate of 2.36$\times 10^{-3}
\rm~counts~s^{-1}$, $\sim 25\%$ of which can attributed to the local
background, as estimated in the \xs-halo region. This fraction
would be reduced by a factor of $\sim 2$, if instead the
background estimated in the 6$^{\prime\prime}$-18$^{\prime\prime}$ annulus is used. Such a difference in the background subtraction would lead to
some quantitative changes, but hardly affect the qualitative results
presented in the present paper.

All our spectral fits are conducted with the XSPEC package
[\cite{Arnaud1996}; version 12.8.0]; model names of the package are
used unless otherwise noted. In particular, the spectrum of the optically-thin thermal
plasma at a certain temperature is modeled with {\it APEC}, which uses the ATOMDB code
(v2.0.1), assuming collisional ionization equilibrium~\cite{Foster2012}. The metal abundances of the plasma, as well as the foreground
X-ray-absorbing gas, are relative to the interstellar medium (ISM) values given in~\cite{Wilms2000}.
The photoelectric absorption uses the Tuebingen-Boulder model [{\it
  TBABS}; \cite{Wilms2000}] with the cross-sections from
\cite{Verner1996}. An alternative version of the plasma model ({\it
  VAPEC}) is sometimes used to allow for abundance variations of individual elements. Both the interstellar dust scattering ({\it DUSTSC}) and
the CCD pile-up ({\it PILEUP}) effects~\cite{Nowak2012} are also
accounted for. 

\begin{figure} 
\includegraphics[width=4.5in]{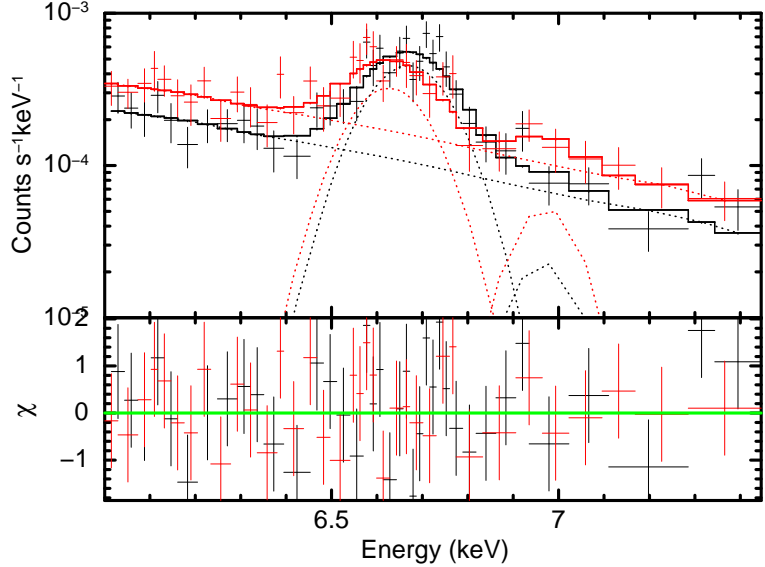}
\caption{
Comparison between the quiescent X-ray spectrum of \xs\ (black; Fig.~\ref{f:fig2}) and the
\xs-halo spectrum (red; extracted from the
2$^{\prime\prime}$-5$^{\prime\prime}$ annulus; Fig.~\ref{f:sfig1}) in
the Fe \ka\ complex region. Two
Gaussian lines are added into the fit to characterize the 6.4~keV and
6.973~keV lines, although the best-fit model fluxes of the former line are
zero for both spectra. the line centroid and EW
of the Fe \ka\ line in the  spectrum of the \xs-halo region, 6.63(6.60 - 6.65)~keV
and 0.29(0.22 - 0.37)~keV, are substantially different from those in the 
spectrum of \xs (Table~\ref{t:line_para}). }
\label{f:sfig3}
\end{figure}

We use a simple 1-T optically-thin thermal plasma to characterize 
the X-ray spectrum of \xs\ in quiescence. To properly
account for distinct emission line features in the spectrum, we 
allow the abundances of the relevant elements
(S, Ar, Ca, and Fe, while all others are tied to C)
to vary independently in the fit. This {\it VAPEC} model
with $kT = 3.5(3.0, 4.0)$~keV gives an
acceptable fit to the spectrum ($\chi^2/n.d.f. =201/216$). But, the
fitted $N_H= 10.1 (9.4, 11.1) \times 10^{22} {\rm~cm^{-2}}$ is far too low to be consistent with that
inferred from the flare spectral analysis. 
Therefore, the model is unlikely to be physical. Nevertheless, we find
that the 1-T plasma with metal abundance set equal to zero gives a
good (simple or ``model-independent'') characterization of  the
spectral continuum. We can then use individual {\it Gaussians} to measure
the emission lines (Fig.~\ref{f:fig2}a). Table~\ref{t:line_para} lists the fitted 
centroid, flux, EW, and identification
for each of the six most significant lines. The table also includes
upper limits to the fluxes and EWs of the diagnostic K$\alpha$  emission lines
from neutral/weakly ionized Fe and H-like Fe.
Similar spectral analysis is also conducted
for the \xs-halo region. The results are compared with those obtained for
\xs\  in Fig.~\ref{f:sfig3}) and in the main text.

\noindent\underline{Implement of the spectral model {\it RIAF}}

While the quiescent X-ray spectrum obtained from the present work gives an unprecedented
diagnostic capability to probe the accretion process of \xs, it is
still not possible to discriminate among many
possible spectral models or their combinations. Therefore, we primarily consider
simple physical models, which capture key
characteristics of the process. In particular, we test the
RIAF solutions [e.g.,~\cite{Narayan1994,Blandford1999,Narayan2000,Quataert2000,Yuan2003,Begelman2012}], by implementing a general spectral model for them. 
In such a solution, the inward radial motion as a
function of the radius $r$ is $v_r =v_o (r_o/r)^{1/2}$, or 
a fixed fraction of the Keplerian orbital velocity  (the subscript $_o$ is used to denote
quantities at the outer radius $r_o$); the accretion rate of the RIAF is $\dot{M} =
\dot{M}_o (r/r_o)^{s}$, where a positive $s$ would indicate a
net mass loss or outflow from the flow. The mass conservation implies that
the density profile is characterized by $n = n_o (r_o/r)^{3/2-s}$.
The temperature is $T = T_o (r_o/r)^\theta$,  increasing with decreasing radius
due to the conversion of the gravitational potential energy into
heat (i.e., $0 < \theta \le 1$). 
 These scaling relations have been largely confirmed by
various hydrodynamic and magneto-hydrodynamic simulations
[e.g.,~\cite{Yuan2012a,Li2013}] as good approximations over broad
radial ranges, although deviations may be expected at the accretion
flow onset region near the Bondi radius
[e.g.,~\cite{Xu2006,Narayan2011}; considerably larger than
our chosen $r_o$] and
in the innermost region ($r_i \lesssim 10^2 r_s$), where the electron temperature 
becomes decoupled from the ion temperature [e.g.,~\cite{Yuan2012a,Shcherbakov2012}].
We calculate the X-ray spectrum of this outer {\it RIAF} model by
integrating the corresponding differential emission measure, $dEM/d{\rm log}(T) \propto
(T_o/T)^{\gamma}$ (where $\gamma=2s/\theta$), together with the emissivity function of the {\sl APEC} plasma
over the temperature range from $T_o$ to $T_i$. The radiation from radii smaller than $r_i$ 
is simply modeled with an independent bremsstrahlung component ({\sl
  BREMSS}) at $T_i$. 

\noindent\underline{{\it RIAF} model fit to the quiescent X-ray spectrum} 

The {\sl RIAF} model constructed above gives
an excellent fit to the spectrum, both globally ($\chi^2/n.d.f. = 187/218$)
and in terms of matching individual lines (Fig.~\ref{f:fig2}b). 
The best-fit model gives $\gamma= 1.9 (1.4, 2.4)$. If $\theta \sim 1$ as typically assumed [e.g.,~\cite{Quataert2000,Begelman2012}], this suggests a very flat density profile of the
flow (i.e., $s \sim 1$), indicating an outflow mass-loss rate that nearly
balances the inflow. 
The fitted metal abundance, 1.5(1.1, 2.1),  and absorption column,
$13.8(12.2, 15.0)\times 10^{22} {\rm~cm^{-2}}$, 
are also consistent with
their expected values. The shape of the bremsstrahlung
spectrum (hence the fit) in the \chandra\ band is 
insensitive to the exact value of $T_{i}$. Its 95\% lower limit is
$\sim 10^8$~K; but 
the best fit value appears to go beyond the upper limit ($\sim 10^9$~K) of
{\sl APEC}. Thus, we fix $T_{i}$  to  this upper limit. 
The best-fit value and the one-sided 95\%  upper limit of $T_o $ are 
$0.96 \times 10^7$~K and $1.5 \times 10^7$~K. No lower limit
is given; the spectral contribution diminishes rapidly with decreasing
temperature, because of the strong soft X-ray absorption along the sight line. 
We use the {\it CFLUX} convolution model to first estimate the
unabsorbed luminosity of \xs\  in the 2-10 keV band as  $3.4
(2.9, 4.3) \times 10^{33} {\rm~erg~s^{-1}}$. 
We then estimate the luminosity of the {\it BREMSS} component to
be  16(5, 23)\% of the total flux of \xs. This fraction is consistent
with the point source contribution 
obtained from the above spatial decomposition. The component 
should have approximately included the relatively small nonthermal
contributions from undetected weak flares and
photons inverse-Compton-scattered into the X-ray band in the very inner region of
the accretion flow [e.g.,~\cite{Yuan2003}]. Therefore, we conclude
that the fraction of the quiescent X-ray emission arises from the
innermost region is about $\sim 20\%$.

\begin{table*}
  \caption{Measurements of individual emission lines}
  \begin{tabular}{lccr}
\\
\hline\hline
Line energy  & flux & EW& ID, expected energy\\
(keV) & ($10^{-6} {\rm~ph~s^{-1}~cm^{-2}}$) & (eV) &  (keV) \\
\hline
2.48 (2.44, 2.52)   &2.5 (1.5, 3.8) & 161 (101, 232)   &S~{\sc XV}, 2.461\\
3.10 (3.03, 3.16)    &0.6 (0.3, 1.0)& 64 (27, 104)     &Ar~{\sc XVII}, 3.140\\
3.35 (3.32, 3.39)    &0.6 (0.3, 0.9)& 72 (37, 109)     &Ar~{\sc XVIII}, 3.32\\
3.91 (3.86, 3.94)    &0.4 (0.2, 0.6)& 63 (31, 96)  &Ca~{\sc XIX}, 3.861\\
6.676 (6.660, 6.691)&1.2 (1.0, 1.4)&691 (584, 846) & Fe~{\sc XXV}, 6.675\\
7.874 (7.737, 8.012)& 0.2 (0.03, 0.4) &181 (91, 417)& Fe~{\sc XXV}, 7.881\\
6.4 (fixed)             & 0 (0, 0.06)      &0 (0, 22) & Fe~{\sc I-XVII}, 6.4\\
6.973 (fixed)          & 0.07 (0, 0.11)      & 0 (0, 42) & Fe~{\sc XXVI}, 6.973\\
\hline
\hline
\end{tabular}
\\

Note: The dispersions of all the Gaussian lines, except for the
strongest one (Fe~{\sc XXV}, 6.675~keV), are fixed at
$10^{-5}$ keV. 
\label{t:line_para}
\end{table*}

\clearpage

\end{document}